\documentclass{article}
\usepackage{graphicx}
\usepackage{amssymb}

\title{String tension and removal of lattice coarsening effects in Monte 
Carlo Renormalization Group}
\author{E.T. Tomboulis\footnote{e-mail: tombouli@physics.ucla.edu}  \ and 
A. Velytsky\footnote{e-mail: vel@physics.ucla.edu}\\
{\em\small 
Department of Physics and Astronomy, UCLA, Los Angeles, CA 90095-1547, USA}}

\begin{document}
\maketitle

\begin{abstract}
We study the computation of the static quark potential 
under decimations in the Monte Carlo Renormalization Group (MCRG).  
Employing a multi-representation plaquette action, we find that 
fine-tuning the decimation prescription so that the MCRG equilibrium 
self-consistency condition is satisfied produces dramatic improvement 
at large distances.  
In particular, lattice coarsening (change of effective lattice spacing  
on action-generated lattices after decimation) is nearly eliminated. 
Failure to correctly tune the decimation, on the other hand, 
produces large coarsening effects, of order $50\%$ or more, consistent 
with those seen in previous studies. We also study rotational 
invariance restoration at short distances, where no particular improvement is 
seen for this action.
\end{abstract}

\section{Introduction}
The construction and study of improved lattice actions has received  
a good deal of attention over the years. In the Wilsonian renormalization 
group (RG), starting from a suitable cut-off action, a 
blocking transformation results into an improved action in the 
sense of being closer to the RG renormalized trajectory.    
In the Monte Carlo Renormalization Group (MCRG) approach some block-spinning 
transformation is applied to a configuration ensemble obtained by Monte Carlo 
simulations. The resulting blocked ensemble is then assumed to be 
Boltzmann-distributed according to some effective action defined on 
the decimated lattice. One, however, does not know at the outset what this 
improved action is. The standard procedure has been to adopt a model for it, 
and then proceed to measure its couplings on the decimated 
configurations by the demon \cite{Creutz:1983ra} or some other method. 

Any effective action model is necessarily restricted to some subspace of 
interactions. This implies that one is always faced with the problem of 
truncation effects, i.e. the issue of whether the space of interactions 
retained in the model is sufficient to adequately describe the 
decimated ensemble at least over some scale regime. 
As demonstrated in our recent work 
\cite{Tomboulis:2007re,Tomboulis:2007rn,Tomboulis:2006si},  
however, there is another issue that must always be separately 
considered. It pertains to the 
self-consistent application of the MCRG method itself: 
the decimated ensemble must belong to the 
equilibrium ensemble of the action model at the couplings obtained 
by measurements on the decimated ensemble. We refer to this requirement as 
the MCRG equilibrium self-consistency condition. 
Correct application of the MCRG method entails checking whether 
this condition is satisfied. Given some choice of
blocking prescription and effective action model, 
one will in fact find that, in general, the 
condition is {\it not} satisfied.

One obvious reason for this may be that the chosen model of the 
blocked effective action simply cannot adequately approximate the 
blocked ensemble over any scale regime. In such a case 
the truncation errors completely 
dominate. There is then nothing to do but include an appropriate  
wider class of interactions in the effective action model.  

Another, more subtle cause for the inconsistency, however, may be in play  
\cite{Tomboulis:2007rn}. 
The `true' blocked action depends on the precise  
choice of decimation prescription.\footnote{Indeed, within the general 
Wilsonian renormalization group framework, the precise  
form of the action resulting from a block-spinning 
transformation cannot be divorced from the choice of blocking  
transformation: different choices for the definition of 
blocked field variables, in terms of the original field variables, will, 
in general, result in different blocked effective actions.} 
It may then be that the assumed effective action model, though 
in principle a quite adequate approximation over some scale regime, has 
not yet been appropriately matched to the chosen decimation procedure 
in the following sense. 
Commonly employed lattice decimation prescriptions typically  
involve adjustable parameters, such as the weights  
of different kinds of staples formed out of the bond variables 
on the  undecimated lattice. In general, the above equilibrium consistency 
condition can be 
satisfied, {\it if at all}, only for particular (range of) values of these 
parameters \cite{Tomboulis:2007rn}. 
In other words, fine-tuning of the decimation prescription is 
needed to match to some appropriate effective action over some scale 
regime.   

This state of affairs, which is generic in the application of MCRG, was 
extensively demonstrated in 
\cite{Tomboulis:2007re,Tomboulis:2007rn,Tomboulis:2006si}  
in decimation studies in $SU(2)$ lattice gauge theory (LGT). 
In these studies, two alternative standard decimation procedures were 
employed: 
Swendsen decimation \cite{Swendsen:1981rb}, and ``double smeared blocking'' 
(DSB) \cite{DeGrand:1994zr} decimation.  
Both decimation prescriptions involve a free parameter $c$, which is the 
weight of staples relative to straight paths in the construction of the 
decimated lattice bond variables out of the original lattice bond variables.
Two different effective action models were explored: the 
multiple-representation single plaquette action (eq. (\ref{eq:eff_act}) 
below)), and the  
fundamental representation plaquette-plus-rectangle  
($1\times 2$ planar loop) action. 

In the case of the multi-representation plaquette action it was found that 
for DSB decimation the parameter $c$ can be sharply fine-tuned so that the 
MCRG equilibrium condition is satisfied. 
The same is true for Swendsen decimations, though there the fine-tuning is 
somewhat less sharp. Overall, DSB decimation turns out to be better suited 
for this  action. By comparing the values of $N\times N$ loop observables 
measured on the decimated ensemble, denoted $W_{N\times N}^{dec}$, to 
those measured on the effective-action-generated ensemble, denoted 
$W_{N\times N}^{gen}$, substantial improvement with increasing length scale 
was found, as expected, at 
the optimal (fine-tuned) $c$ value \cite{Tomboulis:2007rn}. `Improvement' 
here means that the  
difference between $W_{N\times N}^{dec}$ and $W_{N\times N}^{gen}$ 
is reduced (ideally goes to zero), which implies that the blocked ensemble is 
adequately represented by the effective action model. 

In the case of the plaquette-plus-rectangle fundamental representation 
action no $c$ value was found such that DSB-decimated configurations are 
nearly equilibrium configurations of the action. Such a value, however,  
could be found for Swendsen decimations.\footnote{Overall, however, 
this action   
exhibits erratic behavior under variation of $c$, in contrast 
to the smooth, monotonic behavior of the multi-representation plaquette 
action under such variation. This probably indicates instability 
under addition of other terms.}   
Comparison of the (weighted) difference of $W_{N\times N}^{dec}$ and 
$W_{N\times N}^{gen}$ shows that  
the plaquette-plus-rectangle  
works well (nearly vanishing difference) at short distances, but 
gives consistent growth of the difference with increasing $N$ towards 
intermediate distances; it thus appears to fail as an effective action 
at intermediate to long distances. The multi-representation plaquette action 
at the optimal $c$, on the other hand, shows improvement
with increasing length scales, indicating improved efficacy as a longer scale 
effective action.

In \cite{Tomboulis:2007re}, \cite{Tomboulis:2007rn} observables up 
to size $N=8$ (in undecimated lattice units) were investigated. In this 
paper we consider a wider range of scales in measurements of the 
Polyakov line correlator in $SU(2)$ LGT. We obtain the static 
quark-antiquark potential 
which allows us to probe different scales and extract the string tension.  
Comparison of string tensions on the blocked 
and effective-action-generated ensembles allows us then to 
test the extent to which the assumed effective action represents the 
blocked ensemble in terms of a long-distance physical quantity. 
Dramatic improvement is found using the multi-representation plaquette 
action after fine-tuning to the 
$c$ value that satisfies the equilibrium self-consistency requirement. 
For any $c$ values away from this optimal value, on the other hand, large 
deviations, i.e. sizable effective changes of 
scale (coarsening), are found.    
The size of the coarsening effects found away from the optimal $c$ value 
is consistent with that reported in previous studies \cite{Takaishi:1998ag}. 

We also test for rotational invariance improvement at short distances. 
No actual improvement is found with the multi-representation plaquette action. 
This action then appears to provide an effective long distance description 
under DSB decimation at the expense of some distortion at short scales. 
Our findings are further discussed in section \ref{C}.

\section{String tension and coarsening under decimation} 

We start with the $SU(2)$ fundamental representation Wilson action at 
$\beta=2.5$ and perform DSB decimations.  
We take the multi-representation plaquette action 
\begin{equation}
S=\sum_{j=1/2}^{j_N}\beta_j[1-\frac1{d_j}\chi_j(U_p)] \, ,\label{eq:eff_act}
\end{equation}
as our effective action model, which, as reviewed in the previous section, 
appears better 
suited for representing long distance properties of the blocked configurations.
In (\ref{eq:eff_act}) $\chi_j$ denotes the $j$-representation 
character, and $\beta_j$ the corresponding coupling. $U_p$ stands for 
the product of bond variables around a plaquette $p$. 
For simulating the multi-representation action (\ref{eq:eff_act}) we 
use the procedure in \cite{Hasenbusch:2004yq}. To measure couplings we use 
the microcanonical method \cite{Creutz:1983ra}. For 
microcanonical updating and demon measurements we use the algorithm 
in \cite{Hasenbusch:1994ne}. We refer to \cite{Tomboulis:2007rn}, 
\cite{Tomboulis:2007re} for details of our computational procedures and 
numerical simulations.    
The couplings after one DSB decimation with staple weight parameter 
$c$ and scale factor $b=2$ were computed 
for various values of $c$ 
in \cite{Tomboulis:2007rn} (Table V). They are reproduced here for 
some $c$ values in Table \ref{tab:fix_c_dsb}.
\begin{table}[ht]
 \centering
 \begin{tabular}{|c|l|}
 \hline
 $c$ &$\beta_{1/2}, \beta_1, \beta_{3/2}, \beta_2, \ldots$ \\
 \hline \hline
0.060&2.4660(7), -0.3635(11), 0.1242(17),\\ 
      &-0.0475(21), 0.0195(25), -0.0070(24)\\
\hline
 0.065&2.5023(7), -0.3098(12), 0.1057(16),\\
      &-0.0397(16), 0.0145(14), -0.0029(15)\\
 \hline
 0.067&2.5125(7), -0.2832(16), 0.0964(25),\\
       &-0.0367(29), 0.0139(29)\\\hline
 0.068&2.5183(9), -0.2701(13), 0.0916(16),\\
      &-0.0351(17),0.0142(17),-0.0053(20)\\\hline
 0.077&2.5463(11), -0.1167(17), 0.0320(23),\\
      &-0.0055(28)\\\hline 
\end{tabular}
\caption{\label{tab:fix_c_dsb} The couplings of the effective 
action (\ref{eq:eff_act}) 
after DSB decimation starting from the
Wilson action at $\beta=2.5$. $c$ is the staple weight.}
\end{table}

We compute the static quark potential from the Polyakov line 
correlator. The string tension $\sigma$ is then extracted, as usual, 
by fitting the potential $V(r)$ to the expression
\begin{equation}
V(r) = m -{\mu\over r} + \sigma r \, ,\label{qqpot}
\end{equation}
where $r$ is the distance in lattice units.
 
We start on $32^3 \times 12$ lattice with the Wilson action at 
$\beta=2.5$, and obtain the potential $V_0(r)$ and string tension 
$\sigma_0$. 
We next perform a DSB decimation 
with scale factor $b=2$ and several choices of the $c$ parameter. For each 
such $c$ value we measure the potential $V_{dec}$ and 
the string tension $\sigma_{dec}$ 
on the decimated ensemble on the resulting $16^3\times 6$ lattice. 
We then generate configurations for the effective action (\ref{eq:eff_act}) 
at the couplings obtained after the decimation (Table 
\ref{tab:fix_c_dsb}). The potential $V_{gen}$ and the string tension 
$\sigma_{gen}$ are then obtained from this effective-action-generated 
ensemble. For decimated potential measurements we use  
30 replicas of runs, each typical run consisting of up to 10000 sweeps 
with measurement at every 3rd sweep, and  1 heatbath and 
2 overrelaxations per sweep. For the effective action potential measurements 
shorter runs, each typically of 100 sweeps, are used  with measurements 
every 3rd sweep.
For measurements on the original and on the generated ensembles,  
for which the action is known, we use the L\"uscher-Weisz technique; 
whereas, on the decimated ensemble, for which the action is not known, 
we use `naive' straightforward correlation measurements, which 
explains the need for rather longer runs.
Also for this reason, we chose the original lattice at relatively high 
temperature so that the correlator decays slower, allowing us to 
use the `naive' methods on the decimated ensemble. The values of 
the string tensions $\sigma_0$, $\sigma_{dec}$ and $\sigma_{gen}$ 
obtained are listed in Table \ref{tab:string}.
\begin{table}[ht]
\centering
\begin{tabular}{|c||c|c|}
\hline
   &\multicolumn{2}{|c|}{$\sigma_0=$ 0.0313(2)}\\\hline
c    &$\sigma_{dec}$&$\sigma_{gen}$\\\hline
0.060&0.0271(37)&0.0594(12)\\
0.065&0.0284(30)&0.0385(14)\\
0.067&0.0291(49)&0.0346(8)\\
0.068&0.0295(12)&0.0292(9)\\
0.077&0.0285(24)&0.0091(6)*\\
\hline
\end{tabular}
\caption{\label{tab:string}String tensions in original (undecimated) 
lattice units. $\sigma_0$ on $32^3\times 12$, $\sigma_{dec}$, 
$\sigma_{gen}$ on $16^3\times6$, except 
starred entry which is on $16^3\times 8$.}
\end{table}
One sees that the DSB decimated configurations produce correct values 
of the string tension over the range of $c$ values shown. (This is also 
the case for Swendsen decimations, cf. Table I in \cite{Tomboulis:2007rn}.) 
The string tension $\sigma_{gen}$ extracted from the 
effective-action-generated ensemble, however, shows marked 
dependence on the decimation parameter $c$. In fact, for $c=0.077$  
the time-like lattice extension had to be increased to $N_t=8$ to 
keep the generated ensemble in the confined phase. 

\begin{figure}[ht]
 \includegraphics[width=0.9\columnwidth]{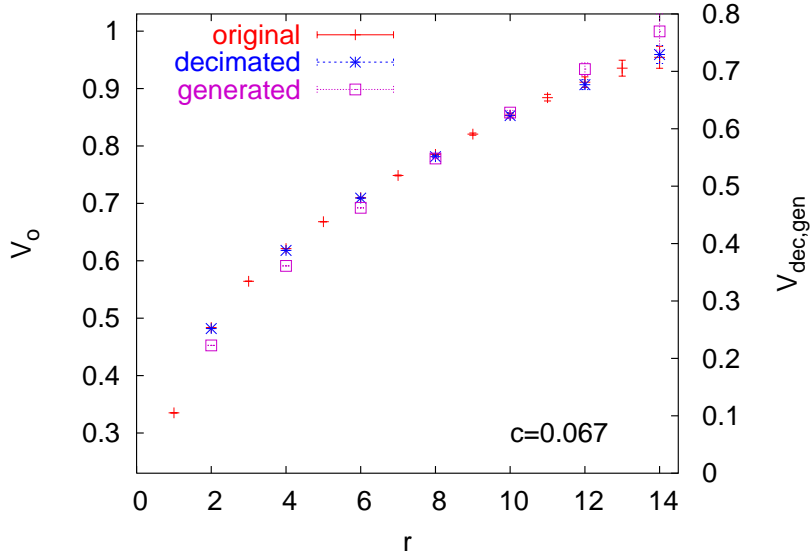}
 \caption{\label{fig:pot1} Static quark potential $V_0$ computed on original 
(undecimated), $V_{dec}$ on DSB decimated ($c=0.067$), and $V_{gen}$ on 
effective-action-generated lattices.}
\end{figure}

In Fig. \ref{fig:pot1} we plot the static quark potential $V_0(r)$ 
obtained on the original (undecimated) $32^3\times 16$ lattice.
To satisfy the MCRG equilibrium self-consistency condition the 
$c$ parameter has to be fine-tuned in the near vicinity of the 
value $c=0.067$ (see \cite{Tomboulis:2007rn}). 
The potentials $V_{dec}$ and $V_{gen}$ 
obtained after DSB decimation for this optimal value of $c$ are 
plotted for comparison also in Fig. \ref{fig:pot1}. The potential
$V_{dec}$ obtained from the decimated ensemble differs from 
$V_0$ in the constant term $m$ representing mass renormalization per 
unit length of the Polyakov lines (external sources). 
Discrepancies in the Coulomb $\mu$ coefficient value (cf. 
\cite{Tomboulis:2007rn}) result in 
some additional distortion, which, however, becomes unimportant with 
increasing $r$. This is as expected; numerical decimation procedures 
typically introduce some short distance distortion. After a shift 
by a constant,  
represented by the relatively shifted l.h.s. and r.h.s. vertical axes 
in Fig. \ref{fig:pot1}, $V_{dec}$ falls for the most part 
nearly on top of $V_0$. This reflects the fact that the string tension 
is well reproduced in $V_{dec}$. 

The important feature of Fig. 
\ref{fig:pot1}, however, 
is that $V_{gen}$, computed at the fined-tuned value 
$c=0.067$, closely tracks $V_{dec}$ and $V_0$ over a wide distance range. 
Note that this occurs without 
the need for any constant shift between $V_{dem}$ and $V_{gen}$, 
which means that the generated ensemble produces nearly the same 
mass renormalization effect $m$ as the decimated effect. 
The near agreement, beyond short distance effects, of $V_{gen}$ with 
$V_{dec}$ and $V_0$ means then that it also reproduces 
the string tension well, as evident from Table \ref{tab:string}. 
The Gaussian difference test for the string tension
values on decimated and generated lattices gives 
$Q=0.27$ (probability that the 
discrepancy is due to chance), an indication of good agreement of data.
To quantify this agreement in physical terms, we consider the actual 
change in scale that resulted from the blocking operations beyond 
the expected scale change by the blocking factor $b=2$. To this end we 
examine the ratios among the lattice spacings in the original, decimated and 
action-generated lattices: 
\begin{equation}
{a_{dec}\over 2a} =\sqrt{\sigma_{dec} \over \sigma_0} \, , \qquad 
{ a_{gen} \over a_{dec}} =\sqrt{\sigma_{gen} \over \sigma_{dec} }\, , \qquad 
{a_{gen}\over 2a} =\sqrt{\sigma_{gen} \over \sigma_0}  \,, \label{eq:ratios}
\end{equation}
where $a$ denotes the lattice spacing on the original (undecimated) lattice. 
If the decimation procedure and the identification of the effective 
action were exact, these ratios would all be equal to one. Deviations 
from unity, which are commonly referred to as `coarsening' errors,  
amount to a change of lattice scale in addition to that by the 
blocking factor $b$ (equal to $2$ here). At our fine-tuned value 
$c=0.067$, one has $a_{gen}/a_{dec}=1.09(9)$,  
$a_{gen}/2a=1.051(13)$, and $a_{dec}/2a= 0.96(8)$. (Here errors are
computed using standard error propagation.) 

Remarkably, then, once the decimation procedure is fine-tuned to satisfy the 
MCRG equilibrium self-consistency condition with the multi-representation 
plaquette action, there is virtually no coarsening effect on the blocked 
lattice.   
 
\begin{figure}[ht]
\includegraphics[width=0.9\columnwidth]{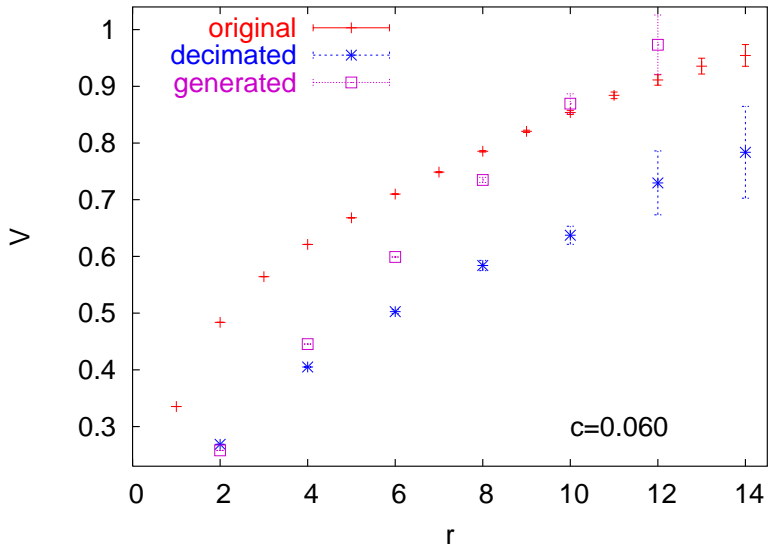}
\caption{\label{fig:pot2} Static quark potential $V_0$ computed on original 
(undecimated), $V_{dec}$ on DSB decimated ($c=0.060$) and $V_{gen}$ on 
effective-action-generated lattices.}
\end{figure}

This is in sharp contrast to what happens at other values of $c$, i.e. 
when the condition is not satisfied. The values for the ratios 
(\ref{eq:ratios}) for different values of $c$ are displayed in Table 
\ref{tab:ratios}. 
\begin{table}[ht]
\centering
\begin{tabular}{|c||c|c|c|}
\hline 
c    &$a_{dec}/2a $&$ a_{gen}/ a_{dec}$&$ a_{gen} / 2a$ \\\hline
0.060& 0.93(6)&1.48(10)  & 1.378(15)\\
0.065& 0.95(5)&1.16(7) & 1.109(20)\\
0.067&0.96(8)&1.09(9)& 1.051(13)\\
0.068& 0.97(2)&0.99(3)&0.966(15)\\
0.077& 0.95(4)&0.57(3)  & 0.539(18)\\
\hline
\end{tabular}
\caption{\label{tab:ratios} Lattice spacings ratios among original, 
decimated and effective-action-generated lattices.}
\end{table}
Even relatively small deviations outside a small window 
inside the interval $0.0065 < c \lessapprox 0.0068$ result in sizable 
coarsening effects. Thus, at $c=0.060$ one has $a_{gen}/a_{dec}= 1.481$, 
i.e. a coarsening effect of about $50\%$ of the action-generated 
lattice compared to the decimated lattice. For illustrative purposes, 
we also plot in Fig. \ref{fig:pot2} the static quark potentials 
$V_0$, $V_{dec}$ and 
$V_{gen}$ at $c=0.060$. The contrast with Fig. \ref{fig:pot1} is manifest. 
Such change-of-scale effects of order $40-50 \%$, and corresponding 
deviations in the blocked potentials as those seen in Fig. \ref{fig:pot2} 
are typical of previous MCRG studies fixing the decimation parameters 
on an ad-hoc basis. For 
the `classical' value $c=0.077$, which has been used before, for example, 
the coarsening effect is $\sim 75\%$!

\section{Rotational invariance restoration}
\begin{figure}[ht]
 \includegraphics[width=0.9\columnwidth]{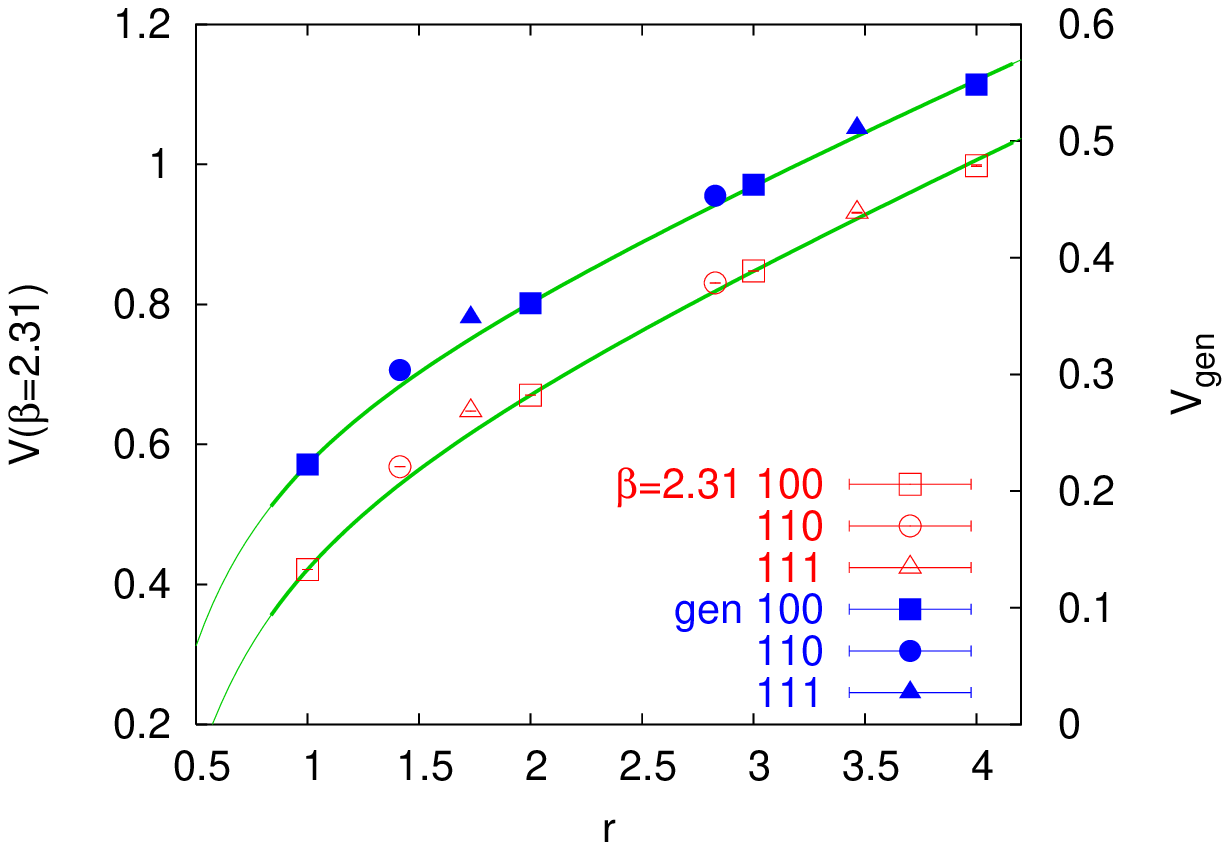}
\caption{\label{fig:pot-off} Static quark-antiquark potential on 
the effective-action-generated  and on $\beta=2.31$ Wilson lattices; on and 
off axis separations.}
\end{figure}

Our decimations have a blocking factor equal to $2$. The Wilson action 
on a lattice of spacing $2a$ would have a string
tension $\sigma=(2a/a)^2 \sigma_0$, i.e. four times the original string 
tension. Therefore, we expect $\sigma\sim0.1252$.
This is very close to the string tension for the Wilson action at 
$\beta=2.31$, which is $\sigma=0.1230(14)$. 
We compare departures from rotational symmetry on this Wilson action lattice
and on the effective-action-generated lattice. 
The time-like Polyakov lines of the correlator intersect a $3$-dimensional 
space-like lattice slice at two points. 
We consider on-axis separation between these two points in the direction 
$100$,  and off-axis separations in the directions $110$ and $111$ 
(in terms of unit vectors in the space-like slice).  
The results are presented in Fig. \ref{fig:pot-off}. 

We quantify the amount of rotational invariance violation by 
\cite{deForcrand:1999bi}: 
\begin{equation}
\delta_V^2=\sum_{off}\left(
\frac{V_{off}(r)-V_{on}(r)}{V_{off}(r)\delta V_{off}(r)}\right)^2/
\sum_{off}\frac1{\delta V(r)^2}. \label{eq:deltaV} 
\end{equation}

We get\footnote{The first four off-axis points, counting from shortest 
distance, in Fig. \ref{fig:pot-off} were included in the sums in 
(\ref{eq:deltaV}).}
$\delta_V=0.045$ for Wilson, and $\delta_V=0.047$ for the effective 
action. They are comparable - there is no improvement at short distance.

\section{Summary and Conclusions}\label{C}

In the MCRG method one has to assume a model for the effective action on 
the blocked lattice. 
Consistent application of the MCRG method then involves checking whether 
the blocked ensemble belongs to the equilibrium ensemble generated from 
the effective action model. In general this test will fail. 
This may indicate that the action model has been restricted to a set of 
interactions that do not adequately approximate the true action 
representing the blocked ensemble, i.e. truncation effects are paramount. 
But, as was pointed out in \cite{Tomboulis:2007re}, \cite{Tomboulis:2007rn}, 
another possible cause for this failure may be that 
the decimation procedure and the assumed effective action have not 
been properly matched. This means that the action may in fact be an adequate 
approximation in some scale regime, but any adjustable parameters that 
enter in the specification of the decimation prescription  have not 
been properly tuned so that the equilibrium self-consistency condition 
is satisfied. 

In this paper we explored this effect by computing  the static quark 
potential and extracting the string tension on the original, the 
decimated and the effective-action-generated ensembles.  
Discrepancies between the string tensions from the different ensembles 
amount to a change in scale in addition to the rescaling from a lattice of 
spacing $a$ to a decimated lattice of spacing $2a$ (for a blocking 
by a factor of $2$). Such discrepancies are referred to as lattice 
coarsening errors.  

Employing DSB decimations with the multi-representation plaquette action 
model we found that fine-tuning the DSB parameter $c$ so that 
the MCRG equilibrium self-consistency condition is fulfilled has a rather 
dramatic effect: lattice coarsening is largely eliminated.  
On the other hand, even small departures from this value of $c$ 
result into sizable deviations between the potentials on the different 
ensembles and correspondingly sizable lattice coarsening effects. 
Coarsenings of $50\%$ or more are typical. One can conclude that for 
the multi-representation plaquette action such 
coarsening arises almost entirely from non-equilibrium rather than actual 
truncation effects.

Since any effective action model will not be quite exact, some truncation 
errors, of size generally varying with distance scale, are always present.  
We tested the multi-representation plaquette action for 
rotational invariance at short distances. No improvement in rotational 
invariance restoration at short distances was observed compared to the Wilson 
action. 

In conclusion, the $SU(2)$ multi-representation plaquette action 
provides a good long-distance representation of the ensemble obtained by DSB 
decimation, at least as far as observables like the static quark potential 
are concerned. For improvement at short distances it has to be augmented, 
presumably by including loops larger than the plaquette. 
It would clearly be very worthwhile to carry out such a  
program in the case of $SU(3)$.   

More generally, a main conclusion of this study is that, 
no matter what its form is, 
no correct assessment of an adopted effective action can be obtained 
without properly  tuning the family of decimations being employed to it. 
The purpose of this tuning must be to satisfy, to the 
extent possible, the MCRG equilibrium self-consistency condition. 
Failure to do this can completely obscure the actual 
efficacy of the action to represent the blocked ensemble and the origin 
of the errors involved.

\section*{Acknowledgments}
We thank Academic Technology Services (UCLA) 
for computer support. This work was in part supported by 
NSF-PHY-0555693.

\bibliographystyle{JHEP}
\bibliography{avref}

\end{document}